\documentclass[12pt]{article}

\usepackage{graphicx}
\usepackage{epstopdf}

\def\onehalf{{\textstyle \frac12}}
\def\ii{{\rm i}}
\def\dd{{\rm d}}
\def\rr{{\bf r}}

\def\ssr#1{{\scriptscriptstyle\rm #1}}
\def\of#1{{{\scriptstyle(}#1{\scriptstyle)}}}
\def\oof#1{{{\scriptscriptstyle(}#1{\scriptscriptstyle)}}}
\def\abs#1{{{\scriptstyle|}#1{\scriptstyle|}}}
\def\aabs#1{{{\scriptscriptstyle|}#1{\scriptscriptstyle|}}}
\def\parder#1#2{\frac{\partial #1}{\partial #2}}
\def\totder#1#2{\frac{\dd #1}{\dd #2}}

\def\tsty#1#2{{\textstyle\frac{#1}{#2}}}

\def\matdos#1#2#3#4{\left(\matrix{\displaystyle{#1}&
			\displaystyle{#2}{}_{\mathstrut}
		\cr \displaystyle{#3}&\displaystyle{#4}\cr} \right)}

\def\jour#1#2#3#4{{\it #1{}} {\bf #2}, #3 (#4)}
\def\lab#1{\label{eq:#1}}
\def\rf#1{(\ref{eq:#1})}
\def\Lie#1{\hbox{\sf #1}}

\newcommand{\be}{\begin{equation}}
\newcommand{\ee}{\end{equation}}
\newcommand{\bea}{\begin{eqnarray}}
\newcommand{\eea}{\end{eqnarray}}

\begin{document}

\begin{center}

{\Large\bf Quantum superintegrable Zernike system} 

\bigskip
        George S.\ Pogosyan,\footnote{Departamento de Matem\'aticas, 
	Centro Universitario de Ciencias Exactas e Ingenier\'ias, 
	Universidad de Guadalajara, M\'exico; Yerevan State University, 
	Yerevan, Armenia; and Joint Institute for Nuclear Research, 
	Dubna, Russian Federation.} Cristina Salto-Alegre,\footnote{Posgrado
        en Ciencias F\'isicas, Instituto de Ciencias F\'isicas-UNAM.}\\
		 Kurt Bernardo Wolf,\footnote{Instituto de Ciencias F\'isicas, 
		 Universidad Nacional Aut\'onoma de M\'exico, Cuernavaca.} 
		 and Alexander Yakhno\footnote{Departamento de 
	Matem\'aticas, Centro Universitario de Ciencias Exactas e 
	Ingenier\'ias, Universidad de Guadalajara, M\'exico.}

\vskip1cm


\vskip1cm

\end{center}

\begin{abstract}
        We consider the differential equation that Zernike proposed
        to classify aberrations of wavefronts in a circular pupil,
        whose value at the boundary can be nonzero.
        On this account the quantum Zernike system, where that differential
        equation is seen as a Schr\"odinger equation with a potential, is special 
        in that it has a potential and boundary condition that are not standard 
        in quantum mechanics. We project the disk on a half-sphere
        and there we find that, in addition to polar coordinates,
        this system separates in two additional coordinate systems
        (non-orthogonal on the pupil disk), which lead to Schr\"odinger-type 
        equations with P\"oschl-Teller potentials, whose 
        eigen-solutions involve Legendre, Gegenbauer and Jacobi polynomials. 
		This provides new expressions for separated polynomial solutions 
		of the original Zernike system that are real. The operators which provide 
		the separation constants are found to participate in a 
		superintegrable cubic Higgs algebra.
\end{abstract}

\vskip1cm

\section{Introduction: the Zernike operator}  \label{sec:one}

The differential operator and eigenvalue equation of 
Zernike \cite{Zernike34} are
\be 
        \hat Z^{(\alpha,\beta)}\Psi\of\rr := \Big(\nabla^2 + 
                \alpha(\rr\cdot\nabla)^2 
                + \beta\,\rr\cdot\nabla \Big)
                        \Psi\of\rr = -E\, \Psi\of\rr,
                        \lab{Zernikeq}
\ee
for real parameters $\alpha$ and $\beta$. In order to describe 
the shape of scalar optical wavefields constrained by 
a unit circular exit pupil, and such that at its
boundary $\abs\rr=1$ the wavefields have constant absolute
value $c=|\Psi\of\rr|_{\aabs\rr=1}$, Zernike found that
for the two-dimensional case, the operator \rf{Zernikeq}
can be self-adjoint under the inner product over the pupil
disk, only when the two parameters have the values 
$\alpha_\ssr{Z}=-1$ and $\beta_\ssr{Z}=-2$, as we show in 
Sect.\ \ref{sec:two}. 

This system and its solutions have many
important properties which have been analyzed 
thoroughly in several optical and mathematical papers
\cite{Bhatia-Wolf,Born,Myrick,Kintner,Wunsche,Shakibaei,Ismail}.
Yet it seems that the symmetries obtained when this
system is projected from the disk on the half-sphere,
have not been yet elucidated up to now. 

The Zernike differential equation \rf{Zernikeq} can evidently 
be separated and solved in polar coordinates $(r,\phi)$. 
As was shown in Ref.\ \cite{PWY}, the classical counterpart 
of this equation describes a system which is separable in 
polar and elliptic coordinates plus, when 
projected on the manifold of a sphere or hyperboloid, 
displays separability in other three and six orthogonal 
coordinate systems respectively. 
In Sect.\ \ref{sec:three} we solve the separated polar and radial 
equations, the former yielding circular harmonics, and the 
latter hypergeometric polynomials that match those of 
Zernike \cite{Zernike34}. The quest for higher symmetries
starts in Sect.\ \ref{sec:four}, where we map the disk 
on a half-sphere with coinciding boundaries. This step
is crucial because it allows the orthogonal coordinates
on the sphere to map onto non-orthogonal coordinates on 
the disk, where the Zernike equation also separates, and
where the separation constants provide extra integrals
of motion.

In Section \ref{sec:five} we introduce three coordinate
systems on the sphere, whose $\vartheta=0$ poles point
along the $z$-, $x$- and $y$-axes. The first returns
essentially the solutions of the previous sections, while the 
other two yield solutions in terms of products of a Legendre and a Gegenbauer
polynomial. In Section \ref{sec:six} the operators that provide the 
separation constants are organized through their commutators
into the nonlinear cubic Higgs superintegrable algebra \cite{Higgs,KKMP}. 
The concluding Sect.\ \ref{sec:seven} recapitulates the
construction and adds some further remarks on Zernike-type
systems.


\section{Boundary conditions and restrictions}  \label{sec:two}

	As we mentioned in the Introduction, the Hilbert space 
of square-integrable functions $f\of\rr\in {\cal L}^2({\cal D})$, 
on the unit disk ${\cal D}:=\{\abs\rr\le1\}$ is determined by 
the inner product
\be 
	(f,g)_{\cal D}
	:= \int_{\cal D}\dd^2\rr\,f\of\rr^*g\of\rr
	=\int_0^1 \!r\,\dd r\int_{-\pi}^\pi \!\dd\phi\, f(r,\phi)^*g(r,\phi), 
			\lab{Z-Hilbert-space}
\ee
where the asterisk indicates complex conjugation, and where 
the functions are required to satisfy the boundary 
value $|f(1,\phi)|={}$constant. In this space, the Zernike operator 
\rf{Zernikeq} is required to be self-adjoint, namely,
\be 
	(f,\hat Z^{(\alpha,\beta)}g)_{\cal D}
		=(\hat Z^{(\alpha,\beta)}f,g)_{\cal D}.
		\lab{self-adjoint}
\ee

Written out in polar coordinates and separated in three 
summands, this operator is
\be 
	\hat Z^{(\alpha,\beta)} = \hat Z_2^{(\alpha)}
		+ \hat Z_1^{(\alpha,\beta)} + \hat Z_\phi,
			\lab{Zernikeab}
\ee
where
\be
	\hat Z_2^{(\alpha)}:=(1+\alpha r^2)\partial^2_r, \quad
	\hat Z_1^{(\alpha,\beta)}:=\bigg( \frac1r
		+ (\alpha+\beta)r\bigg)\partial_r, \quad
	\hat Z_\phi:= \frac1{r^2}\partial^2_\phi.
				\lab{Three-Z}
\ee
On each summand the integral \rf{Z-Hilbert-space} will 
be performed by parts yielding boundary terms. The
last term we can immediately integrate by parts 
over $\phi$, yielding
\be 
	(f,\hat Z_\phi g)_{\cal D} = (\hat Z_\phi f, g)_{\cal D}
		+\int_0^1\frac{\dd r}{r}\bigg(
		(f^*\partial_\phi g - g\partial_\phi f^*)
			\Big|_{\phi=-\pi}^\pi\bigg).
		\lab{intZphi}
\ee
The last term will evidently vanish when the functions
are single-valued over the disk, so we can consider
\be 
	f(r,\phi)=f_m\of{r} \frac{e^{\ii m\phi}}{\sqrt{2\pi}},
	 \lab{sep-var}
\ee	 
with any integer $m$. Let us continue indicating by $f\of{r},\,g\of{r}$,
functions of the radius $r$ alone, suppressing their
index $m$, and obviating the integral over $\phi$ 
in \rf{Z-Hilbert-space} that will yield unity.

The first-order differential term $\hat Z_1^{(\alpha,\beta)}$
in \rf{Three-Z} will be now integrated by parts over $r|_0^1$, 
giving a left-over integral and a boundary term,
\be 
	(f,\hat Z_1 g)_{r}= -(\hat Z_1 f,g)_{r}
		-2(\alpha+\beta)\!\int_0^1\!r\,\dd r\,f^*g
		+\Big(1+(\alpha+\beta)r^2\Big)f^*g\Big|_0^1.
			\lab{intZ1}
\ee
Proceeding similarly with the second-order differential term 
$\hat Z_2^{(\alpha)}$, we obtain
\be 
	\begin{array}{rcl} 
	(f,\hat Z_2 g)_{r}&=& \displaystyle (\hat Z_2 f,g)_{r} 	+\int_0^1\!\dd r\,
		\Big(2(1+3\alpha r^2)(\partial_rf^*)g + 6\alpha r f^*g\Big)\\
		&&\displaystyle {}+\bigg(r(1+\alpha r^2)
			\Big(f^*\partial_r g - (\partial_r f^*)g\Big) 
			-(1+3\alpha r^2)f^*g\bigg)\bigg|_0^1. \end{array}
				\lab{intZ2}
\ee
Summing \rf{intZ1} and \rf{intZ2} yields
\be 
	\begin{array}{l}
	\Big(f,(\hat Z_2{+}\hat Z_1) g\Big)_{\!r}
		=\Big( (\hat Z_2{-}\hat Z_1)f, g\Big)_{\!r}\\ \displaystyle
	{\quad}+2\int_0^1\!r\,\dd r\,\Big((2\alpha-\beta)f^*g + 
		(1/r + 3\alpha r)(\partial_rf^*)g\Big) \\ 
	\displaystyle{\quad}+\bigg(r(1+\alpha r^2)\Big(f^*\partial_rg
		-(\partial_rf^*)g\Big)+ r^2(\beta-2\alpha)f^*g\bigg)
			\bigg|_0^1. \end{array} \lab{Z1plusZ1}
\ee
The boundary term is zero at $r=0$; for $r=1$ and generally 
nonzero values for $f(1)$, $g(1)$ or their derivatives, the first summand
vanishes when $\alpha=-1$, and then the coefficient of second 
summand will also vanish when $\beta=2\alpha=-2$; 
for these values of $\alpha$ and $\beta$, the remaining integral 
term in the right-hand side of \rf{Z1plusZ1} will then be 
$2\int_0^1 r\,\dd r\,(1/r-3r)(\partial_rf^*)g
=2(\hat Z_1^{(-1,-2)}f, g)_{r}$, as can be seen from \rf{Three-Z}.
The last term $\hat Z_\phi$ in \rf{intZphi} is independently self-adjoint,
so it follows that the Zernike operator $\hat Z^{(-1,-2)}$ satisfies 
the required self-adjointness condition \rf{self-adjoint}.

Given the form of the angular part of the Zernike differential 
operator $\hat Z_\phi$ in \rf{Three-Z}, its eigenfunctions
being $\sim e^{\ii m\phi}$ for all integers 
$m\in\{0,\pm1,\pm2,\ldots\}$,  we may separate 
the solutions $\Psi\of\rr$ of \rf{Zernikeq} as
\be 
	\Psi(r,\phi) := R^\oof{m}\of{r}\frac{e^{\ii m\phi}}{\sqrt{2\pi}},
		\lab{sep-var1}
\ee
turning the Zernike equation \rf{Zernikeq} into an ordinary 
differential equation for the radial factor $R^\oof{m}\of{r}$, 
\be
	r^2(1-r^2)\totder{^2 R^\oof{m}\of{r}}{r^2}  
	+r(1-3r^2)\totder{R^\oof{m}\of{r}}{r} 
	 - m^2 R^\oof{m}\of{r} = -Er^2\,R^\oof{m}\of{r}, 
	\lab{ZernikeRad}
\ee
where the values of $E$ will be determined by the
square-integrable solutions that can be normalized
as $R^\oof{m}(1)={}$constant. 	


\section{The Zernike basis of functions on the disk}  \label{sec:three}

The radial differential equation of Zernike \rf{ZernikeRad} is
of hypergeometric type. Writing $R^\oof{m}\of{r}= r^mF(r^2)$,
the factor $F\of{z}$ is solution of the hypergeometric equation
\cite[Eq.\ 9.151]{GR}, 
\be 
	z(1{-}z)F''+\Big( (m{+}1)-(m{+}2)z\Big)F' 
		- \tsty14\Big(m(m{+}2)-E\Big)F=0,
			\lab{hyperg}
\ee
which has one solution of the form ${}_2F_{\!1}(a,b;c;z)$, 
with parameters
\be 
	a=\onehalf(m{+}1)\pm\onehalf\sqrt{E{+}1}, \quad
	b=\onehalf(m{+}1)\mp\onehalf\sqrt{E{+}1}, \quad
	c=m+1.   
	\lab{abc}
\ee
Since $m$ is integer and $c$ must be positive, the
absolute value $\abs{m}$ should be understood for
$\surd m^2$ in \rf{ZernikeRad}. Also, since $c=a+b$, 
the solution will be logarithmically singular at $z=r^2=1$ 
unless the hypergeometric series terminates and is 
a polynomial. This occurs when we write $E:=n(n+2)$ 
and ask $n-\abs{m}$ to be an even non-negative integer, 
thus defining the {\it radial\/} quantum number
\be 
	n_r:=\onehalf(n-\abs{m})\in\{0,1,2,\ldots\},
		\lab{radqn}
\ee	
and the energy $E$ in \rf{Zernikeq} is then given by the 
{\it principal\/} quantum number $n$,
\be 
	 E = n(n+2), \quad n=2n_r+\abs{m}\in \{0,1,2,\ldots\}.
			\lab{radqn}
\ee 

Hence, the square integrable solutions to the radial 
Zernike equation \rf{ZernikeRad} in the interval 
$r \in [0, 1]$ are of the form
\bea
	R_n^m(r) &:=& A_{n,m} \, r^{\aabs{m}} \, 
		{}_2F_{\!1}(-n_r,\, n_r + \abs{m} +1;\ \abs{m} + 1;\, r^2)
			\lab{Zernikesol}\\
			&=& A_{n,m}\,\bigg({n_r+\abs{m}\atop \abs{m}}\bigg)^{-1}
			 r^{\aabs{m}} P^{(\aabs{m},0)}_{n_r}(1{-}2r^2),
			\lab{Zernikeorig}
\eea
where $A_{n,m}$ is a constant and we recognize the identity of 
the hypergeometric with Jacobi polynomials of degree $n_r$ 
in $(1-2r^2)$ \cite[Eq.\ 8.962.1]{GR}.

Zernike's original requirement \cite[Eq.\ (22)]{Zernike34} was
that $R_n^m(1)=1$, leading to choose the constant $A_{n,m}$  
in \rf{Zernikesol}--\rf{Zernikeorig} given by a sign and 
binomial coefficient, so that  
$A^\ssr{Zernike}_{n,m}:=(-1)^{n_r}\Big({n_r+\aabs{m}\atop\aabs{m}}\Big)$
defines his disk polynomials as
\be 
	Z^m_n(r,\phi) := R^m_n(r)\,\bigg\{ \begin{array}{rl}
		\cos m\phi, &\hbox{for } m\ge0,\\
		\sin m\phi, &\hbox{for } m<0.\end{array} 
			\lab{Zernkdef}
\ee

In the present paper we prefer to attend the `quantum-mechanical' 
normalization of the disk functions, using the orthogonality 
of the Jacobi polynomials over $r\in[0,1]$ in the form 
\cite[Eq.\ 7.391]{GR} 
\be 
	\int_0^1 r\,\dd r\,\Big|r^\aabs{m} 
		P^{(\aabs{m},0)}_{n_r}(1{-}2r^2)\Big|^2
		=\frac1{2(n+1)}.  \lab{intJac}
\ee
Since $\int_{-\pi}^\pi \dd\phi=2\pi$, we adopt the 
normalization constant for the disk functions as
$A_{n,m}=\sqrt{2(n{+}1)/2\pi}$ in \rf{sep-var1}, 
so they are  
\be 
	\Psi_n^m(r,\phi):= 	(-1)^{n_r}\sqrt{\frac{n+1}{\pi}}\,
	r^\aabs{m} P^{(\aabs{m},0)}_{n_r}(1{-}2r^2)\,
			e^{\ii m\phi},  \lab{ourZernik}
\ee
with $n=2n_r+\abs{m}$. At the center of the disk
$\Psi_n^m(0,\phi)=0$ for $m\neq0$, while (for $n$ even)
$\Psi_n^0(0,\phi)=\sqrt{(n{+}1)/\pi}$, and
$\Psi_0^0(r,\phi)=1/\surd\pi$. At the circle boundary $r=1$,
\be 
	\Psi_n^m(1,\phi) = \tsty18(n+\abs{m})(n+\abs{m}-2)
		\sqrt{\frac{n+1}{\pi}}\,e^{\ii m\phi}.
				\lab{valueatcenter}
\ee
These wavefunctions satisfy the orthonormality relation 
\be 
	(\Psi_n^m,\Psi_{n'}^{m'})_{\cal D} =	
	\int_{\cal D} \dd^2\rr\, 
	\Psi_n^m\of\rr^*\,\Psi_{n'}^{m'}\of\rr
		=\delta_{n,n'}\,\delta_{m,m'},
			\lab{orthoZernik}
\ee
and are solutions to the quantum Zernike Hamiltonian equation
\be 
	-\hat Z\Psi_n^m\of\rr := \Big({-\nabla}^2 
                +(\rr\cdot\nabla)^2 
                +2\,\rr\cdot\nabla \Big)
             \Psi_n^m\of\rr = n(n+2)\, \Psi_n^m\of\rr.
                        \lab{QZernikeq}
\ee 

Density plots of the Zernike disk polynomials are 
ubiquitous in the literature and on the web, so
we need not reproduce here the real and imaginary
parts of $\Psi_n^m(r,\phi)$ in \rf{ourZernik}. 
Below we shall display the new disk polynomials
associated with separating coordinates different
from the polar ones.


\section{Finding additional constants of motion}  \label{sec:four}

For a fixed value of energy $E=n(n+2)$ given by the 
principal quantum number $n$ in \rf{radqn}, there 
is a range of radial and azimutal quantum
numbers $n_r$ and $m$ that sum to $n=2n_r+\abs{m}$.
The degeneracy in $\pm m$ stems from the \Lie{SO($2$)}
rotational symmetry of the disk $\cal D$ generated by
the angular momentum operator
\be 
	\hat L :=  x \partial_y - y \partial_x.
		\lab{angmomop}
\ee
But there is also a larger degeneracy between those
two quantum numbers, present in the multiplets
\be 
	m\in\{n,\,n-2,\,\ldots\,{-n}\},
		\lab{degm}
\ee
that keep $n-m$ as even integers, and which indicates 
an \Lie{SU($2$)} symmetry and extra integrals of
motion that we proceed to find. These must be of 
second degree in momentum, and would imply that
other systems of separating coordinates exist. As is
well known in two-dimensional flat space, the Helmholtz 
and Schr\"odinger equations allow separation of 
variables in four orthogonal systems, namely in Cartesian,
polar, parabolic, and elliptic coordinates \cite{Miller}.
A simple analysis of the Zernike equation \rf{Zernikeq} on
the unit disk $\cal D$ shows that only the polar system
evinces this separation, so the question of
existence of additional integrals of motion and of
separating coordinates is open. Below we shall solve
this problem by finding two integrals of the
motion in addition to $\hat L$ in \rf{angmomop}, which
is the only obvious one.

Consider again the Zernike operator \rf{Zernikeq} with
the values of $\alpha=-1$ and $\beta=-2$ that we saw
in Sect.\ \ref{sec:two} to allow its self-adjointness 
on the unit disk $\cal D$, written in Cartesian coordinates,
\be 
	\hat Z := (1-x^2)\partial_{xx} - 2 x y \partial_{xy} 
		+  (1-y^2)\partial_{yy} - 3(x\partial_x + y\partial_y).	
			\lab{ZerniCart}
\ee
Now we perform the similarity transformation 
\be 
	\widehat W := A \hat Z A^{-1},\quad A\of{r}
			:=(1-x^2-y^2)^{1/4}= (1-r^2)^{1/4},
	\lab{Aofr}
\ee	
to obtain the new operator
\be 
	\begin{array}{rcl}
	\widehat W &=& (1-x^2)\partial_{xx} - 2 x y \partial_{xy} +  
		(1-y^2)\partial_{yy} - 2(x\partial_x + y\partial_y)\\[5pt]
		&&{}+\tsty14(1-x^2-y^2)^{-1} + \tsty34. \end{array}
			\lab{Wernike}
\ee

\begin{figure}[t]
\centering
\includegraphics[scale=0.60]{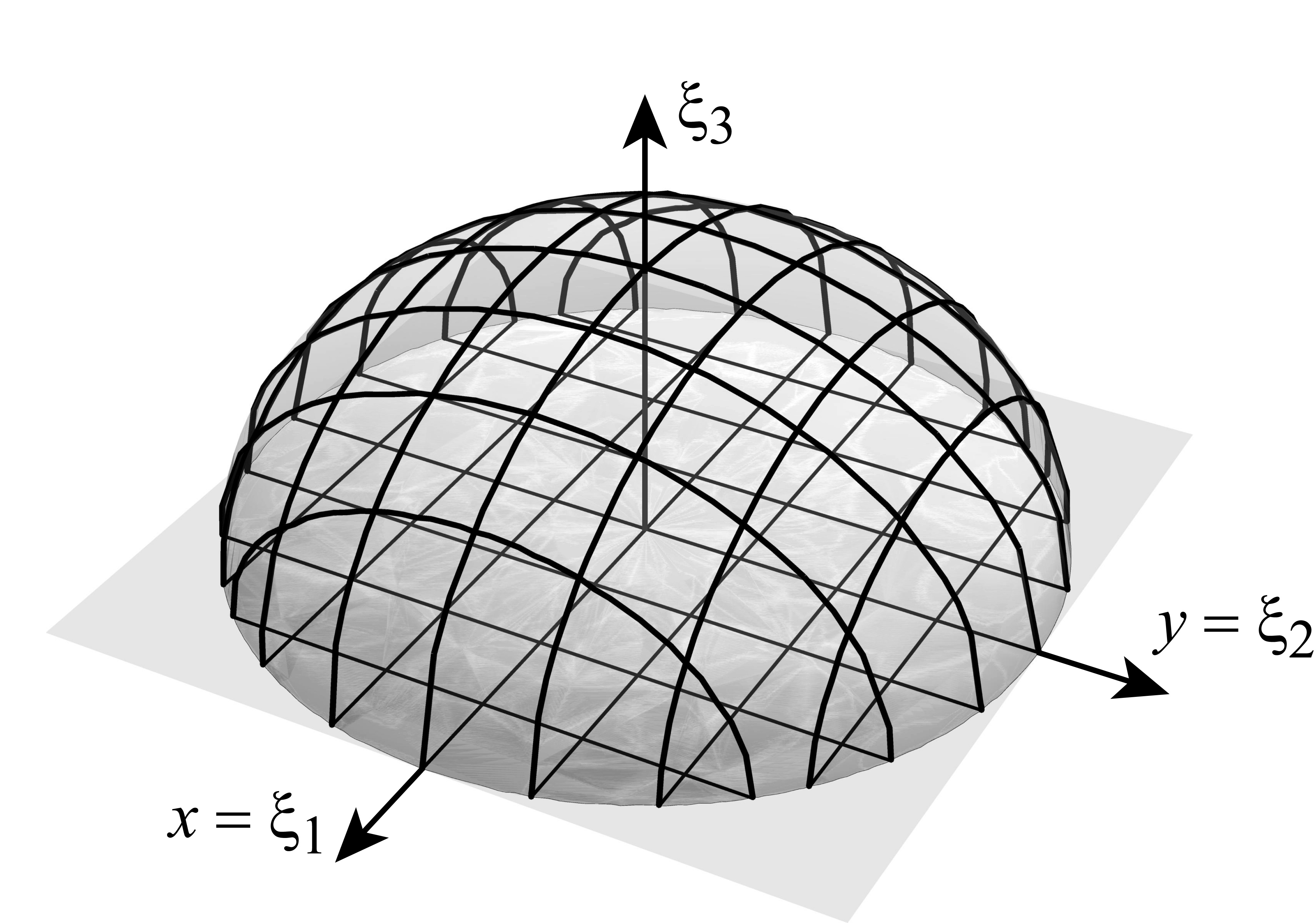}
\caption[]{Map of the unit disk $\cal D$ on the unit 
upper hemisphere ${\cal H}_+$ through the orthogonal projection
\rf{Euc3} of Cartesian coordinates.}
\label{fig:hemisphere}
\end{figure} 

As in the classical system \cite{PWY}, we shall map 
the unit disk $\cal D$ on the upper hemisphere ${\cal H}_+$,
$\xi_1^2+\xi_2^2+\xi_3^2=1$, $\xi_3\ge0$, 
embedded in a three-dimensional Euclidean space of coordinates 
$\{\xi_i\}_{i=1}^3$, using the orthogonal (or `vertical')
projection as shown in Fig.\ \ref{fig:hemisphere},
\be 
	\xi_1 = x, \quad \xi_2 = y, \quad  
		\xi_3 = \sqrt{1-x^2-y^2}, \lab{Euc3}
\ee
where $\xi_1^2+\xi_2^2=r^2$, while the partial derivatives map on 
$\partial_i:=\partial/\partial\xi_i$ as
\be 
	\partial_x=\partial_1-\frac{\xi_1}{\xi_3} \partial_3,\qquad
	\partial_y=\partial_2-\frac{\xi_2}{\xi_3} \partial_3.
		\lab{mappart}
\ee
The second-order operator $\widehat W$ in \rf{Wernike}, with
$\partial_{ij} = \partial^2/\partial_{\xi_i}\partial_{\xi_j}$, thus becomes
\bea 
	\widehat W &=& (\xi_2^2 + \xi_3^2) \partial_{11} 
		+ (\xi_1^2 + \xi_3^2) \partial_{22} 
		+ (\xi_1^2 + \xi_2^2) \partial_{33} \nonumber\\ 
	&&{}- 2\xi_1 \xi_2\partial_{12} 
		-2\xi_1 \xi_3\partial_{13} -2 \xi_2 \xi_3\partial_{23} 
		-2\xi_1\partial_1 - 2\xi_2\partial_2 - 2\xi_3\partial_3\lab{OpW1}\\ 
	&&{}+ \frac{\xi_1^2 + \xi_2^2}{4\xi_3^2} + 1 \nonumber\\
	&=& \Delta_\ssr{LB} + \frac{\xi_1^2+\xi_2^2}{4\xi_3^2} + 1, \lab{OpW2}
\eea
where we have introduced the Laplace-Beltrami operator 
on the two-dimensional unit sphere
\be 	
	\Delta_\ssr{LB}:= \hat L_1^2 + \hat L_2^2 + \hat L_3^2, \lab{LapBel}
\ee
and where $\{L_i\}_{i=1}^3$ are the generators of an \Lie{SO($3$)} Lie algebra, 
\be
	\hat L_1 := \xi_3 \partial_2 - \xi_2 \partial_3, \quad
	\hat L_2 := \xi_1 \partial_3 - \xi_3 \partial_1, \quad 
	\hat L_3 := \xi_2 \partial_1 - \xi_1 \partial_2. \lab{ThreeL}
\ee

While the metric on the disk $\cal D$ is diagonal and distance is 
$\dd s^2=\dd x^2 +\dd y^2$, the metric on the surface
of the half-sphere ${\cal H}_+$ of $|\vec\xi\,|=1$, is
\be 
	{\bf g}=\matdos{1{+}\xi^2_1/\xi^2_3}{\xi_1\xi_2/\xi^2_3
	}{\xi_1\xi_2/\xi^2_3}{1{+}\xi^2_2/\xi^2_3},\quad
	g:=\det{\bf g}= \frac1{\xi^2_3}=\frac1{1-(\xi^2_1{+}\xi^2_2)},
	   \lab{metric}
\ee
so that distance is $\dd s^2=\sum_{i,j=1}^2 g_{i,j}\dd\xi_i\,\dd\xi_j$,
and the surface elements on ${\cal H}_+$ and $\cal D$ are related by
\be 
	\dd^2V(\vec\xi\,)= \sqrt{g}\,\dd\xi_1\,\dd\xi_2 
		= \frac{\dd\xi_1\,\dd\xi_2}{\xi_3}
		=\frac{\dd x\,\dd y}{\sqrt{1-(x^2{+}y^2)}} 
		= \frac{\dd^2{\bf r}}{\sqrt{1-r^2}}.
		\lab{surf-elem}
\ee
This clearly shows that the measure on ${\cal H}_+$ grows 
when $\xi_3\to0$ ($r\to1$) so that its vertical projection 
on the disk remains constant up to the boundary. 

As a result, the quantum Zernike Hamiltonian  equation \rf{QZernikeq} 
on the unit disk $\cal D$, written in terms of $\widehat W$, transforms 
to a quantum Schr\"odinger equation on the unit upper half-sphere 
${\cal H}_+$ for wavefunctions $\Upsilon_n^m(\vec\xi\,)$ of the form
\be 
	 \Big( {-\Delta}_\ssr{LB} 
		-\omega^2\frac{\xi_1^2+\xi_2^2}{\xi_3^2}\Big)\Upsilon_n^m(\vec\xi\,) 
		= (E+1)\Upsilon_n^m(\vec\xi\,),
		\lab{Higgs-osc}
\ee
which corresponds to a form of {\it repulsive\/} oscillator potential,
\be 
	V_\ssr{\!R}(\vec\xi\,):= -\onehalf w^2 \frac{\xi_1^2+\xi_2^2}{\xi_3^2}
	 = -\onehalf w^2 \frac{r^2}{1-r^2},
		\lab{VHiggs}
\ee
that generalizes the superintegrable Higgs attractive 
oscillator \cite{Higgs,GPS,KMP}, to a repusive one
with {\it negative\/} coupling constant $-\onehalf w^2$, 
whose wavefunctions are
\be 
	\Upsilon_n^m(\vec\xi\,):=A\of{r}\,\Psi_n^m\of\rr
		=(1-r^2)^{1/4}\Psi_n^m(r,\phi)
		\lab{Hwavef}
\ee
where $\rr$ is $(\xi_1,\xi_2)$ or $(r,\phi)$, and with energy eigenvalues 
\be 
	{\cal E}:=\onehalf(E+1)=\onehalf(n+1)^2,\quad 
		n\in\{0,1,2,\ldots\}. \lab{energies}
\ee
Because $A(1)=0$, the wavefunctions $\Upsilon_n^m(\vec\xi\,)$ 
in \rf{Hwavef} vanish on the boundary $\xi_3=0$
of ${\cal H}_+$, while at the `top pole' $\xi_3=1$, $r=0$,
they have the values found for $\Psi_n^m\of\rr$ after 
\rf{ourZernik}. 

From the orthonormality relation between the wavefunctions
$\Psi_n^m\of\rr$ when integrated over the disk $\cal D$
in \rf{orthoZernik} for the inner product 
$(\circ ,\circ)_{\cal D}$, under the proper inner product on
the half-sphere ${\cal H}_+$ due to \rf{surf-elem} and 
\rf{Hwavef}, the corresponding orthonormality of the 
wavefunctions $\Upsilon_n^m(\vec\xi\,)$ is
\be
	(\Upsilon_n^m,\Upsilon_{n'}^{m'})_{{\cal H}_+}
		:=\int_{{\cal H}_+}\dd^2V(\vec\xi\,)\,
		\Upsilon_n^m(\vec\xi\,)^*\,\Upsilon_{n'}^{m'}(\vec\xi\,)
	 = (\Psi_n^m,\Psi_{n'}^{m'})_{\cal D} 
	=\delta_{n,n'}\delta_{m,m'}.
	\lab{inn-prod-Y}
\ee


\section{Solution to the Schr\"odinger equation \rf{Higgs-osc}}
			\label{sec:five}

The key to analyze the Zernike system in new light has
been to map the unit disk $\cal D$ on the half-sphere 
${\cal H}_+$. It is on this manifold that one can
introduce in a natural way other coordinate systems.
Indeed, the Higgs repulsive oscillator system \rf{Higgs-osc} can 
be separated in four systems of coordinates: three mutually 
orthogonal spherical  systems of coordinates \cite{PSW1}, namely
\bea
        &&\hskip-30pt\hbox{System I:} \lab{Syst-I} \\
                &&\hskip-30pt\xi_1 = \sin\vartheta\cos\varphi,\quad 
                \xi_2=\sin\vartheta\sin\varphi,\quad
                \xi_3=\cos\vartheta,\qquad
                \vartheta|_0^{\pi/2},\ \varphi|_0^{2\pi},
                \nonumber\\
        &&\hskip-30pt\hbox{System II:} \lab{Syst-II} \\
                &&\hskip-30pt\xi_1 = \cos\vartheta',\quad 
                \xi_2=\sin\vartheta'\cos\varphi',\quad
                \xi_3=\sin\vartheta'\sin\varphi',\qquad
                \vartheta'|_0^\pi,\ \varphi'|_0^\pi,
                \nonumber\\
        &&\hskip-30pt\hbox{System III:} \lab{Syst-III} \\
                &&\hskip-30pt\xi_1 = \sin\vartheta''\sin\varphi'',\quad 
                \xi_2=\cos\vartheta'',\quad
                \xi_3=\sin\vartheta''\cos\varphi''.\qquad
                \vartheta''|_0^\pi,\ \varphi''|_{-\pi/2}^{\pi/2},
                \nonumber
\eea
and also the {\it elliptic\/} coordinate system.

\begin{figure}[t]
\centering
\includegraphics[scale=0.60]{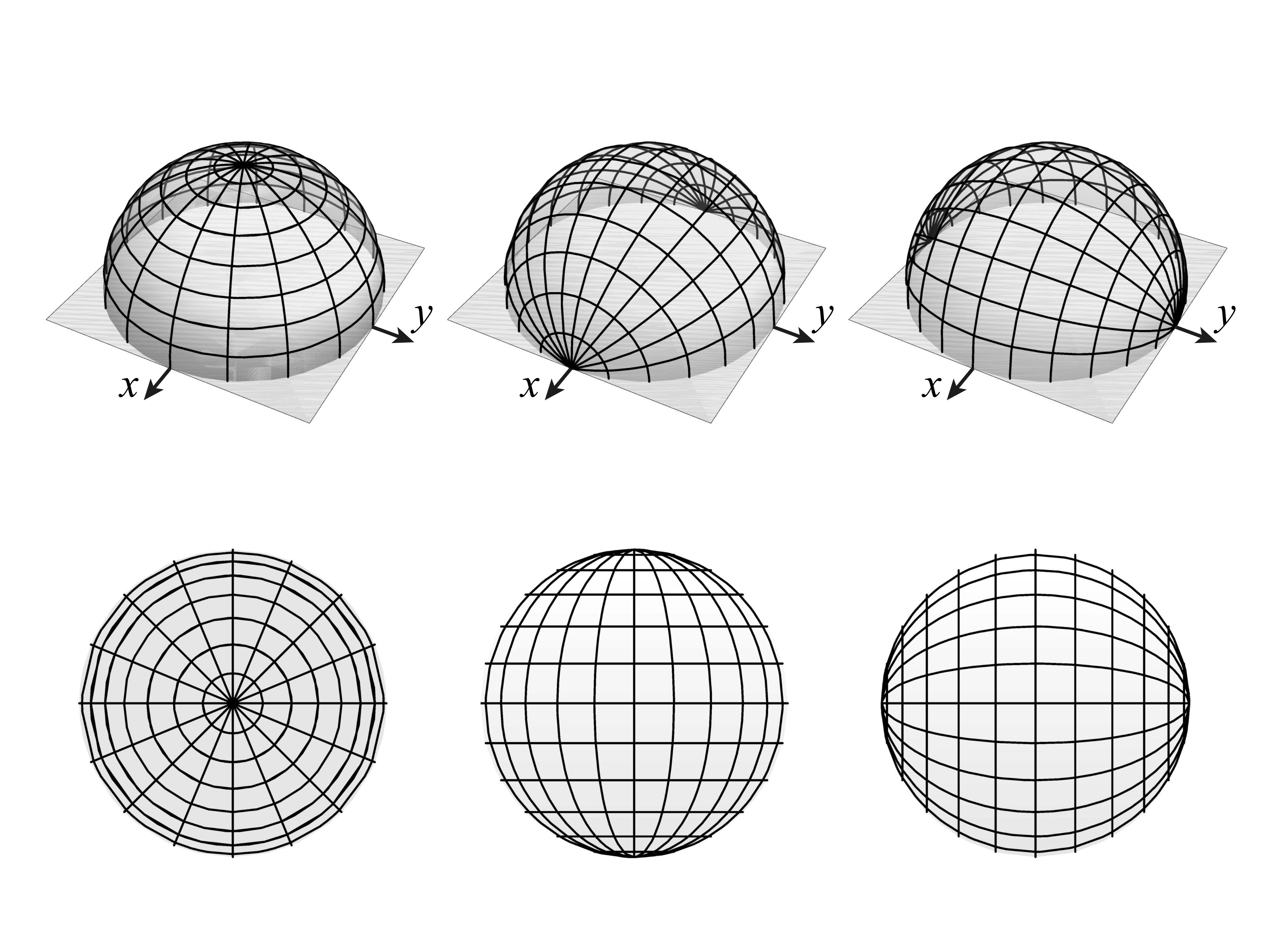}
\caption[]{The coordinate systems \rf{Syst-I}--\rf{Syst-III}.
{\it Top row\/}: on the half-sphere ${\cal H}_+$, where
the $\vartheta=0$ pole is directed along the vertical
$z$-axis, and on the $x$- and $y$-axes. {\it Bottom row\/}: The
same coordinate systems after projection over the disk $\cal D$.}
\label{fig:non-orthog}
\end{figure}

Restricting our consideration in this paper only 
to the above three spherical systems, we 
now examine the form of the potential present in each.
In Fig.\ \ref{fig:non-orthog} we show the three coordinate
systems \rf{Syst-I}--\rf{Syst-III} on the sphere and on
the projected disk, on which the solutions in this section
will separate, and to appreciate that the latter two 
coordinate systems, while they are orthogonal over the sphere, 
they are non-orthogonal over the disk. Normally such 
coordinates are not considered when examining separability
on a flat space.

\subsection{The system I in \rf{Syst-I}}

In the spherical coordinate system $(\vartheta,\varphi)$ of 
\rf{Syst-I} the repulsive oscillator potential 
\rf{VHiggs} takes the form
\be 
	V_\ssr{\!R}(\vartheta) = - \frac{\xi_1^2+\xi_2^2}{8\xi_3^2}
			= -\tsty{1}{8} \tan^2\vartheta,  \lab{VI}
\ee
and the corresponding Schr\"odinger equation \rf{Higgs-osc} 
has the form
\be 
	\frac{1}{\sin\vartheta}\frac{\partial}{\partial \vartheta}
	\sin\vartheta\frac{\partial \Upsilon^\ssr{I}(\vartheta,\varphi)}{\partial \vartheta}
	+\frac{1}{\sin^2\theta}\frac{\partial^2 \Upsilon^\ssr{I}(\vartheta,\varphi)}{\partial\varphi^2}
	+ (2 {\cal E} + \tsty{1}{4}\tan^2\vartheta) \Upsilon^\ssr{I}(\vartheta,\varphi) = 0.
\ee
We now separate the wave function according to the coordinates 
$(\vartheta,\varphi)$,
\be 
	\Upsilon^\ssr{I}(\vartheta, \varphi) =
		\frac{Z^\ssr{I}(\vartheta)}{\sqrt{\sin\vartheta}}\,
			\frac{e^{\ii m \varphi}}{\sqrt{2\pi}},\qquad
			m \in \{0,\pm1,\pm2,\ldots\}, \lab{YI}
\ee
so we come to find $Z^\ssr{I}(\vartheta)$ as the 
solution of a `singular' P\"oschl-Teller-type equation,
\be 
	\frac{\dd^2 Z^\ssr{I}(\vartheta)}{\dd\vartheta^2}+\bigg(2{\cal E}
	-\frac{m^2-\frac14}{\sin^2\vartheta} 
	+ \frac{1}{4\cos^2\vartheta} \bigg) Z^\ssr{I}(\vartheta) = 0. \lab{ZI}
\ee

\begin{figure}[t]
\centering
\includegraphics[scale=0.5]{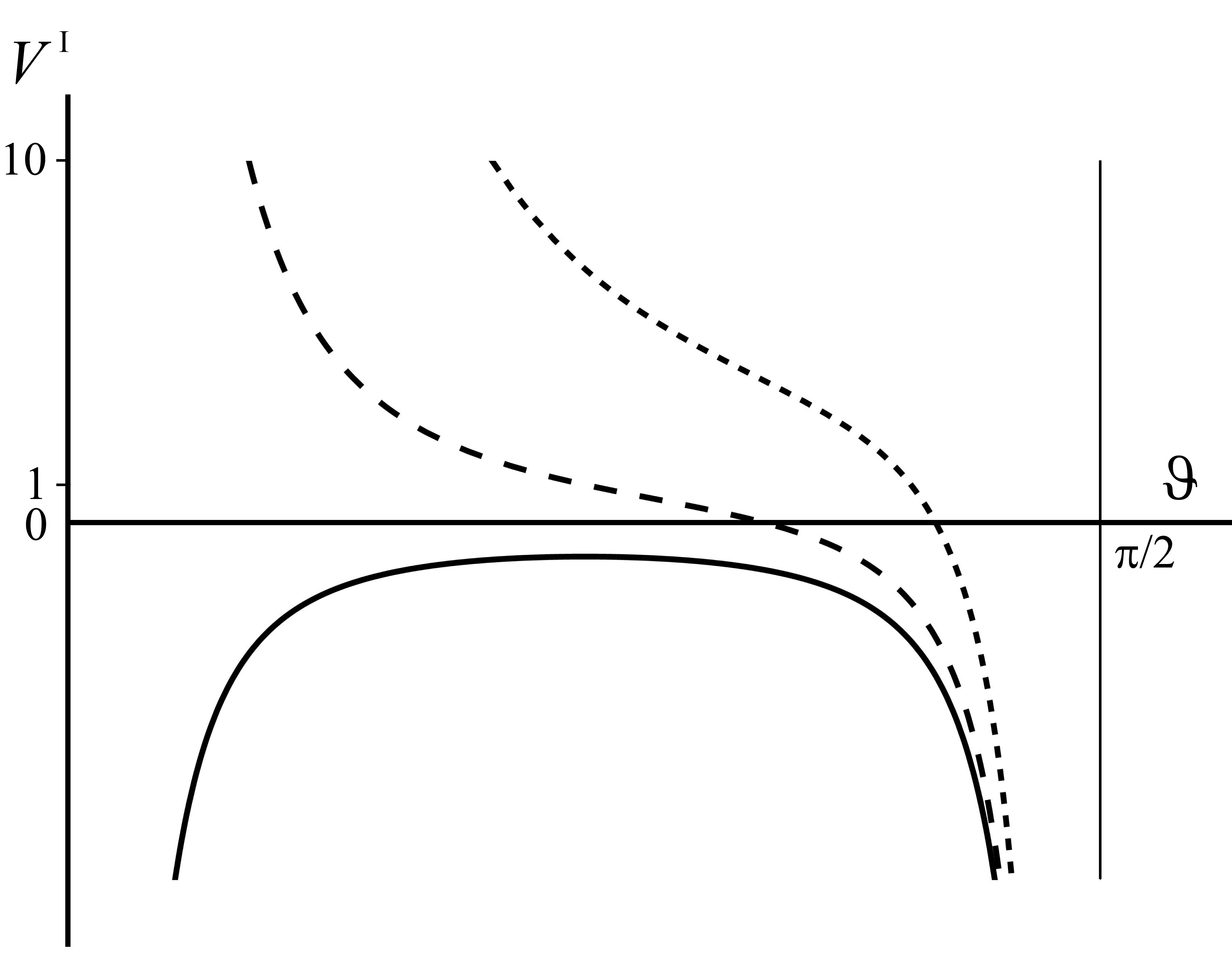}
\caption[]{Effective potential $V_\ssr{\!eff}^\ssr{I}(\vartheta) = 
(m^2-1/4)/\sin^2\vartheta - 1/4 \cos^2\vartheta$ in \rf{VI}, for 
$\vartheta\in(0,\pi/2)$ and values 
$m=0$ (continuous line), $m=1$ (dashed), and $m=2$ (dotted).}
\label{fig:EffPotI}
\end{figure}

This equation describes the one-dimensional quantum wavefield in
the effective potential
\be 
	V_\ssr{\!eff}^\ssr{I} (\vartheta) = \frac{m^2-\frac14}{\sin^2\vartheta} 
	- \frac{1}{4\cos^2\vartheta},  \lab{VI}
\ee
shown in Fig.\ \ref{fig:EffPotI}, which contains a strong repulsive 
singularity at $\vartheta = 0$ (for $m \neq 0$) and a weak 
attractive singularity at $\vartheta = \onehalf\pi$, where
we choose the self-adjoint extension with positive spectrum;
when $m=0$ both singularities are weak and we follow the same choice. 
Such singularities of the P\"oschl-Teller potentials have been 
considered in \cite{Frank}, and appear also in the coupling
Clebsch-Gordan coefficients of two lower-bound `discrete' representations of
the Lorentz algebra \Lie{so($2,1$)} \cite{Basu-KBW}. 

While in the general P\"oschl-Teller potential (on a finite interval) 
one may have both positive and negative energies, we will have
solutions of the Schr\"odinger equation whose potential
\rf{VI} has only positive energy eigenvalues.
Our task now is to 
find the square-integrable solutions of Eq.\ \rf{ZI} 
that satisfy the boundary conditions of vanishing at
the singularities $\vartheta=0$ and $\onehalf\pi$ of \rf{VI},
\be 
	Z^\ssr{I}(0) = 0, \qquad Z^\ssr{I}(\onehalf\pi)= 0, 
		\lab{0-PT}
\ee
with the additional requirement that at the boundary,
\be
	Z^\ssr{I}(\vartheta)/\sqrt{\cos\vartheta}\Big|_{\vartheta=\pi/2} 
		= \hbox{constant}\neq0.
		\lab{extrbound}
\ee
This requirement embodies the factor 
$A(r)=(1-r^2)^{1/4}=\surd\cos\vartheta$ introduced 
in \rf{Aofr}, and allows $\Upsilon^\ssr{I}(\vartheta, \varphi)$ 
in \rf{YI} to be nonzero at the boundary $r=1$.

For the boundary conditions \rf{0-PT}, 
the energy spectrum of ${\cal E}$ in \rf{ZI} is positive and 
discrete,  namely
\be 
	{\cal E} = \onehalf(n+1)^2, \qquad n\in\{0,1,2,\ldots\},
				\lab{1-PT}
\ee
as  determined by $E$ in \rf{radqn}.
To prove this proposition we replace in \rf{ZI} the new variable 
$s := \sin^2\vartheta$ and substitute 
\be 
	Z^\ssr{I}(\vartheta) = s^{(\abs{m}+\frac12)/2} (1-s)^{1/4} f\of{s},
		\lab{Zsf}
\ee
where $f\of{s}$ now satisfies
\be 
	s(1-s) f^{''} + \Big((\abs{m}+1) - s(\abs{m}+2)\Big) f^{'} - 
		\tsty14\Big((\abs{m}+1)^2 -  2{\cal E}\Big) f = 0.
			\lab{eqforf}
\ee
The solution of this equation that is regular at $s=0$ is a
hypergeometric function, 
\be 
	f\of{s} = C \,  {}_2F_{\!1}
			\Big(\onehalf(\abs{m}{+}1 + \sqrt{2{\cal E}}),\,
			\onehalf(\abs{m}{+}1 - \sqrt{2{\cal E}});
			\abs{m}{+}1; s \Big), 		\lab{f-hypergeom}
\ee
where $C$ is a constant. 
The second solution to \rf{eqforf} diverges logarithmically
at $s=0$, i.e., at $\vartheta=0$ and hence at the center of
the disk $r=0$, so we disregard it.

Still, since the parameters of the hypergeometric function in
\rf{f-hypergeom} again sum as $a+b=c$, its behaviour at $s=1$ 
will also diverge logarithmically, as was the case in \rf{abc} 
for polar coordinates of the disk $\cal D$, and nevertheless the
two boundary conditions in \rf{0-PT} are satisfied due to 
\rf{Zsf}. To have solutions $\Upsilon^\ssr{I}$ that can be a nonzero 
constant over the circle $r=1$ the third boundary condition 
\rf{extrbound} must hold, and again this requires 
the hypergeometric series to terminate as a polynomial.
There is thus a subtle difference between quantization
on the disk as performed in Sect.\ \ref{sec:three}, and
quantization on the half-sphere as done here. We must
therefore demand that one of the two first parameters of 
the hypergeometric function in \rf{f-hypergeom} 
be zero or a negative integer, which leads us to define again 
the radial quantum number
\be 
	n_r:=-\onehalf(\abs{m} + 1 - \sqrt{2{\cal E}})\in\{0,1,2,\ldots\},
		\lab{radial-qnumb}
\ee
as we did to find the spectrum in \rf{radqn}, thus
proving the assertion in \rf{1-PT}. We thus define the principal 
and radial quantum numbers related by the angular momentum parameter 
$\abs{m}=\surd m^2$ in \rf{YI} and the P\"oschl-Teller potential 
\rf{VI} by $n = 2n_r + \abs{m}$, and use them them to label the
solutions in \rf{ZI} as $Z^\ssr{I}_{n_r,m} (\vartheta)$. Using
the  boundary condition \rf{extrbound} to determine the appropriate
constant $C$ in \rf{f-hypergeom} we write thus the solution with
the two quantum number labels as 
\bea
	Z^\ssr{I}_{n_r,m} (\vartheta) &=& \sqrt{2(n+1)} \frac{n_r!\,\abs{m}!}{(n_r+\abs{m})!} \, 
			(\sin\vartheta)^{\abs{m}+1/2}\, (\cos\vartheta)^{1/2} \,
					\nonumber\\
		&&{}\times {}_2F_{\!1}(- n_r,\, n_r + \abs{m} +1;\, \abs{m}+1; \sin^2\vartheta)
			\lab{sol-1}\\[5pt]
		&=& \sqrt{2(n+1)} \, (\sin\vartheta)^{\abs{m}+1/2}\,(\cos\vartheta)^{1/2} 
			P_{n_r}^{(\abs{m}, 0)}(\cos 2\vartheta), \lab{sol-2}
\eea
where again $P^{(\alpha, \beta)}_n (u)$ are the Jacobi polynomials,
as was the case in the polar coordinate case \rf{Zernikesol}.
The wave functions $Z^\ssr{I}_{n_rm} (\vartheta)$ in the 
interval $\vartheta\in [0,\onehalf\pi]$ of ${\cal H}_+$ are normalized as
\be 
	\int_{0}^{\pi/2}\!\!\! \dd \vartheta\, 
		Z^\ssr{I}_{n_r, m} (\vartheta)^{*} \,  Z^\ssr{I}_{n'_r, m}(\vartheta) 
		= \delta_{n_r,n'_r}, \lab{ZIintnorm}
\ee
which yields the orthonormalization for the $\Upsilon^m_n(\vartheta,\varphi)$
solution in \rf{inn-prod-Y}. 

Returning from the variables $(\vartheta, \varphi)$ of System I 
in \rf{Syst-I} to the polar coordinates $(r, \phi)$, with 
$r=\sin\vartheta$ and $\phi=\varphi$, as shown in Fig.\ \ref{fig:non-orthog}
(left), taking into account the connection 
between the functions $\Upsilon^\ssr{I}(\vartheta, \phi)$ in \rf{YI} 
and $\Psi(r, \phi)$, and attaching the principal quantum
number label, we obtain the result \rf{ourZernik}.


\subsection{The system II in \rf{Syst-II}}

In the second spherical coordinate \rf{Syst-II}, 
the potential \rf{VHiggs} expressed in the coordinates 
$(\vartheta', \varphi')$, is now 
\be 
	V^\ssr{II}_\ssr{\!eff} 
	= - \frac{1}{8}\bigg(\frac{1}{\sin^2\vartheta' \sin^2\varphi'}-1\bigg).
		\lab{VeffII}
\ee
The corresponding quantum Zernike Hamiltonian equation 
\rf{Higgs-osc} can be separated with the substitution
\be 
	\Upsilon^\ssr{II}(\vartheta', \varphi') = \frac{1}{\sqrt{\sin\vartheta'}}
		S(\vartheta')\,T(\varphi'),  \lab{eigenY}
\ee
so we come to a system of two differential equations
with a separation constant $k$, 
\be 
	\frac{\dd^2 S}{\dd\vartheta^{'2}}+ \bigg( 2 {\cal E}
		-\frac{k^2-\frac{1}{4}}{\sin^2\vartheta'}\bigg) S = 0,
		\qquad
	\frac{\dd^2 T}{\dd\varphi^{'2}} + \bigg(k^2
		+  \frac{1}{4 \sin^2\varphi'}\bigg)T = 0.
			\lab{two-eqs}
\ee
These equations can be put in form where the 
P\"oschl-Teller form is more evident introducing
the new variables $\mu=\onehalf\varphi'$ and
$\nu=\onehalf\vartheta'$, as
\bea
	\totder{^2T(\mu)}{\mu^2}+\bigg(4k^2+\frac1{4\sin^2\mu}
		+\frac1{4\cos^2\mu}\bigg) T(\mu)&=&0, \lab{TT1}\\
	\totder{^2S(\nu)}{\nu^2}+\bigg(4{\cal E}^2
		+\frac{1-4k^2}{4\sin^2\nu}+\frac{1-4k^2}{4\cos^2\mu}\bigg) S(\nu)&=&0. 
		\lab{SS1} 
\eea
The boundary condition at the weak singularities of \rf{TT1}  
were discussed following Eq.\ \rf{ZI}, while those of \rf{SS1}
are even weaker due to the $-4k^2$ summand. Regarding the
extra boundary condition analogue to \rf{extrbound} now is
\be 
	T(\mu)/\surd \cos\mu\Big|_{\mu=\pi/2} = \hbox{constant} \neq0.
		\lab{extrT}
\ee

Solving  these equations we obtain the constant and 
the energies $\cal E$ in \rf{1-PT}
\be
	k = n_1+\onehalf, \quad
		{\cal E} = \onehalf(k+n_2+\onehalf)^2 =
	\onehalf(n_1+n_2+1)^2 = \onehalf(n+1)^2
	\lab{sep-const-k}
\ee
where $n=n_1+n_2$ is the principal quantum number and 
$n_1, n_2 \in \{0,1,2,\ldots\}$, so that the energy 
spectrum is the same as in previous case. 

The solution to both equations \rf{two-eqs} is similar and
the orthonormalized eigenfunctions \rf{eigenY} can be 
written, labelled by the two quantum numbers and separation
constant, as
\be 
	\Upsilon^\ssr{II}_{n_1,n_2}(\vartheta', \varphi') 
		= C_{n_1,n_2}\,\sin^{n_1+\frac12}\vartheta'
		\sin^{\frac12}\varphi'\, 
		C_{n_2}^{n_1+1}(\cos\vartheta')\,P_{n_1}(\cos\varphi'),
				\lab{solIV}
\ee
where
\be 
		C_{n_1,n_2}:= 2^{n_1+\frac12}n_1!
		\sqrt{\frac{(2n_1+1)(n_1+n_2+1)\,n_2!}{2\pi\,(2n_1+n_2+1)!}},
			\lab{solvV}
\ee
and where $C_n^{\gamma}(z)$ and $P_n(z)$ are the Gegenbauer 
and Legendre polynomials of degree $n$ in $z$, respectively. 

\begin{figure}[t]
\centering
\includegraphics[scale=0.15]{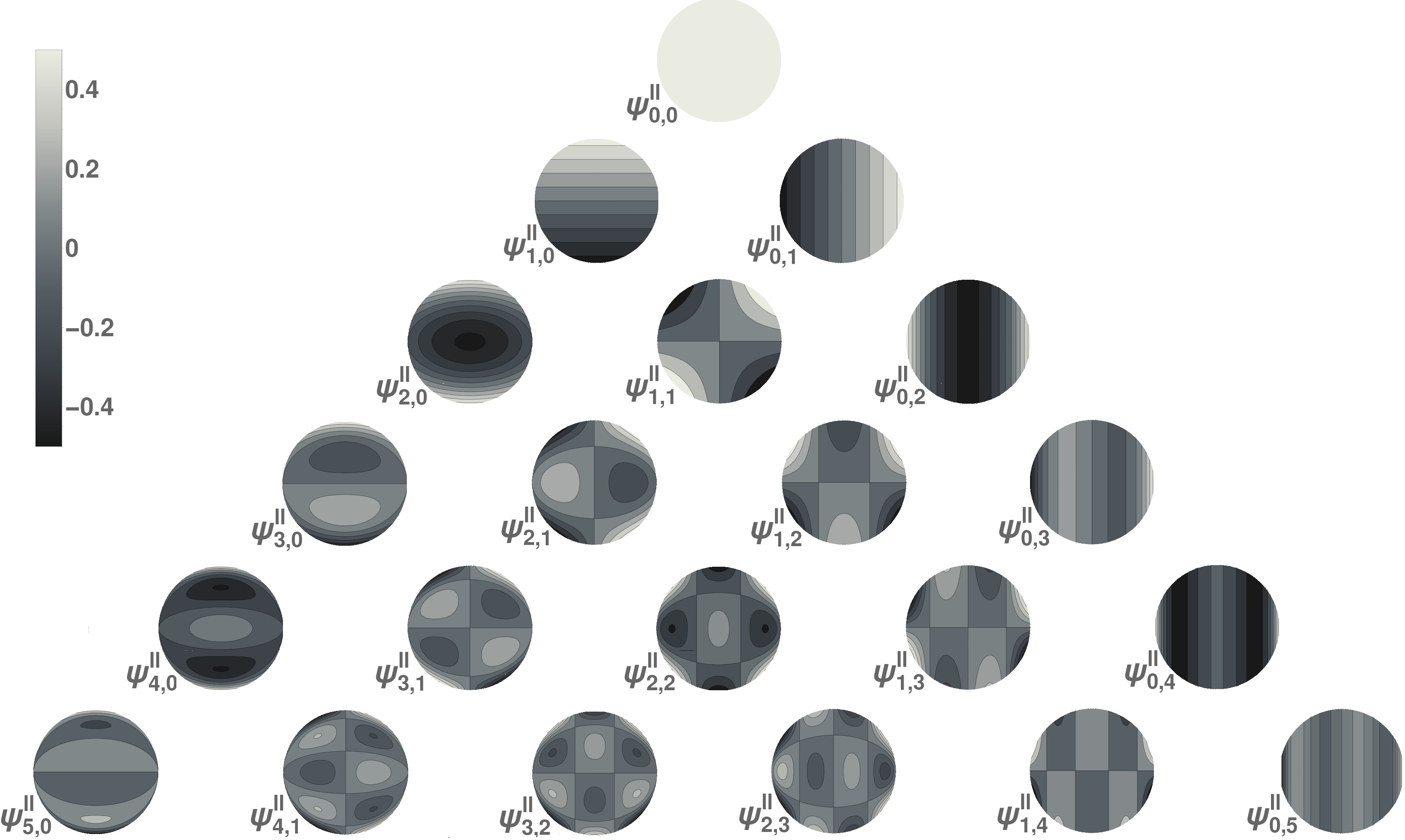}
\caption[]{The new polynomial solutions to the
Zernike quantum system on the disk $\cal D$, 
$\Psi^\ssr{II}_{n_1,n_2}(x,y)$ in \rf{PsiII},
with rows of the same principal quantum
number $n=n_1+n_2$.
There are ten tones of gray between 
contours to emphasize the separating coordinates.}
\label{fig:PsiII-contours}
\end{figure}

We note that the operator that characterizes the 
separation of the solutions in this coordinate system 
involves the operator $\hat L_1$ in \rf{ThreeL}, and is
\be 
	\begin{array}{rcl} 
	\hat J_1 \Upsilon^\ssr{II}_{n_1,n_2}(\vartheta', \varphi')
	 &:=&\displaystyle \bigg(\frac{\partial^2}{\partial \varphi'^2}
		+\frac{1}{4 \sin^2\varphi'} \bigg)
			\Upsilon^\ssr{II}_{n_1,n_2}(\vartheta', \varphi')\\[8pt] 
		&=&\displaystyle\bigg(\hat L_1^2 + \frac{\xi_2^2+\xi_3^2}{4 \xi_3^2}\bigg)
			\Upsilon^\ssr{II}_{n_1,n_2}(\vartheta', \varphi') 
			= - k^2 \Upsilon^\ssr{II}_{n_1,n_2}(\vartheta', \varphi'),
		\end{array}\lab{JsepII}
\ee
where we recall that $k=n_1+\onehalf$. 
Finally, we return to the $(x,y)$ coordinates on the
disk $\cal D$ through $\cos\vartheta'=x$, $\cos\varphi'=y/\sqrt{1-x^2}$, 
to write the wavefunctions as
\be 
	\Psi^\ssr{II}_{n_1,n_2}(x,y) =C_{n_1,n_2}\,(1-x^2)^{n_1/2}\,C^{n_1+1}_{n_2}(x)\,
			P_{n_1}\Big(\frac{y}{\sqrt{1-x^2}}\Big).
				\lab{PsiII}
\ee
In this form it is evident that these solutions are real
and nonzero at the boundary except for isolated points
where the polynomials vanish. In Fig.\ \ref{fig:PsiII-contours}
we provide a density plot for these functions on the disk.

The Zernike differential equation \rf{Zernikeq} was found
rather easily to separate in polar coordinates $(r,\phi)$,
where for the Zernike values $(\alpha,\beta)=(-1,-2)$,
the radial part was \rf{ZernikeRad}. Having here separated its 
solutions by coordinates $(u,v):=(x,\,y/\sqrt{1-x^2})$
that are shown in Fig.\ \ref{fig:non-orthog} (middle),
we see that the solutions can be written as 
$\Psi_{n_1,n_2}(x,y)=U_{n_1,n_2}\of{u} V_{n_1}\of{v}$, 
and the equation written in separated form as follows,
\be 
	\begin{array}{l} \displaystyle
	(u^2{-}1)^2 \parder{^2\Psi}{u^2} 
		+ 3u(u^2{-}1)\parder{\Psi}{u}\\
	\displaystyle{}\quad{} +(1{-}v^2)\parder{^2\Psi}{v^2} 
		- 2v\parder{\Psi}{v} 
		= E(u^2-1)\Psi.   \end{array}  \lab{sepsep}
\ee
The disk $\cal D$ in $(x,y)$ is thus mapped on the
square $\abs{u}\le 1$, $\abs{v}\le1$ where the
coordinates $(u,v)$ are orthogonal.


\subsection{The system III in \rf{Syst-III}}

The same line of reasoning we followed above for Systems I and II, 
apply to the coordinate system III in \rf{Syst-III} for the
coordinates $(\vartheta'', \varphi'')$. There, the potential \rf{VHiggs}
also takes the form of an effective potential also of P\"oschl-Teller type,
\be 
	V^\ssr{III}_\ssr{eff} = - \frac{1}{8}\bigg(\frac{1}
			{\sin^2\vartheta'' \cos^2\varphi''}-1\bigg).
			\lab{VeffIII}
\ee
This potential stems from \rf{VeffII} through the exchange
$\vartheta'\to\vartheta''$ and $\varphi'\to\varphi''+\onehalf\pi$. 
The solution of the Schr\"odinger equation \rf{Higgs-osc} 
in the coordinate system III, now has the separated form,
\be 
	\Upsilon^\ssr{III}_{l_1, l_2}(\vartheta'', \varphi'') =
	C_{l_1,l_2}\,\sin^{l_1+\frac12}\vartheta''	\cos^\frac12\varphi'' 
	\,C^{l_1+1}_{l_2}(\cos \vartheta'')\,P_{l_1}(\sin\varphi''),
	 \lab{new-Upsilon}
\ee
with $l_1, l_2 \in \{0,1,2,\ldots\}$, the same constant 
\rf{solvV} and principal quantum number $n=l_1+l_2$.
 The energy spectrum is also given by $\cal E$ in \rf{1-PT}.

The additional operator that describes the separation of 
solutions in the System III is
\be 
	\begin{array}{rcl} 
	\hat J_2 \Upsilon^\ssr{III}_{l_1, l_2}(\vartheta'', \varphi'')&:=& \displaystyle
	\bigg(\frac{\partial^2}{\partial \varphi''^2}
	+ \frac{1}{4 \cos^2\varphi''}\bigg)
	\Upsilon^\ssr{III}_{l_1, l_2}(\vartheta'', \varphi'')\\[8pt] 
	&=&\displaystyle
	\bigg(\hat L_2^2 + \frac{\xi_1^2+\xi_3^2}{4 \xi_3^2}\bigg) 
		\Upsilon^\ssr{III}_{l_1, l_2}(\vartheta'', \varphi'')
	= - l^2 \Upsilon^\ssr{III}_{l_1, l_2}(\vartheta'', \varphi'')\end{array}
		\lab{JsepIII}
\ee
where $l:=l_1+\onehalf$.
The expression of the wavefunctions \rf{new-Upsilon} in the
original coordinates $(x,y)$ on the disk, using $\cos\vartheta''=y$ 
and $\cos\varphi''= x/\sqrt{1-y^2}$, is
\be 	
	\Psi_{l_1,l_2}(x,y) = C_{l_1,l_2}(1-y^2)^{l_1/2}
		\,P_{l_1}\Big(\frac{x}{\sqrt{1-y^2}}\Big)\,
			C^{l_1+1}_{l_2}(y).  \lab{PsiinIII}
\ee
This coincides with \rf{PsiII} under the 
rotation $x\to y$ and $y\to -x$ which connects
systems II and III. The density plots of 
$\Psi_{l_1,l_2}(x,y)$ are thus identical to
those in Fig.\ \ref{fig:PsiII-contours}, except for
a $\onehalf\pi$ rotation of the disks.


\section{The superintegrable algebra of Zernike} \label{sec:six}

The two operators that determined the constans of motion, 
$\hat J_1$ in \rf{JsepII} and $\hat J_2$ in \rf{JsepIII},
were written in terms of the angular momentum operators
$\hat L_i$ in \rf{ThreeL}. We can add the angular
momentum $\hat L_3$ in \rf{angmomop} and \rf{ThreeL} as 
a third one, and thus have
\be 
	{\hat J_1} = \hat L_1^2  +  \frac{\xi_2^2 + \xi_3^2}{4 \xi_3^2},\quad
	{\hat  J_2} = \hat L_2^2  +   \frac{\xi_1^2 + \xi_3^2}{4 \xi_3^2},\quad
	{\hat  J_3} = \hat L_3, \lab{ThreeJ}
\ee
and thereby write the operator $\widehat W$ in \rf{OpW2} as
\be 
	\widehat W = {\hat J_1} + {\hat J_2} + {\hat J_3^2} + \onehalf. \lab{WinJ}
\ee

To complete this algebra, we construct a third 
linearly independent operator out of the commutator
of the previous two, 
\be 
		\hat S_1 = {\hat  J_3}, \qquad
	\hat S_2 = \hat  J_1-\hat  J_2, \qquad
	\hat S_3 = [\hat S_1,\hat S_2]
\ee
which now satisfy the the following relations:
\be 
	\hat S_3=2\{\hat L_1 , \hat L_2\}_\ssr{\!+} -  \frac{\xi_1\xi_2}{\xi_3^2}, \quad 
		[\hat S_3, \hat S_1]= 4 \hat S_2, \quad
			[\hat S_3, \hat S_2] = 8 \hat S_1^3 - 8 \widehat W \hat S_1,
\ee
where $\{\,,\,\}_\ssr{\!+}$ is the anticommutator. 
Thus, the operators $\hat S_1, \hat S_2, \hat S_3$ a 
generate nonlinear algebra, called the cubic or Higgs algebra \cite{Higgs}.

To write the three operators that commute with Zernike operator $\hat Z$
in the original configuration space $(x,y)$, we must undo the similarity 
transformation in \rf{Wernike} for the symmetry operators 
$\hat K_i = A^{-1} \hat S_i A$, thus obtaining three constants 
of the motion,
\bea
	\hat K_1 &=& y\partial_x - x\partial_y, \lab{KK1}\\
		\hat K_2 &=& -(1-x^2-y^2) 
			(\partial_{xx} - \partial_{yy}) + 2x\partial_x -
	2y\partial_y, \lab{KK2}\\
	\hat K_3 &=& -4 (1-x^2-y^2) \partial_{xy} +
		 4y\partial_x + 4x\partial_y,\lab{KK3}
\eea
which close into the algebra  
\be 
	[\hat K_1,\hat K_2] = \hat K_3, \quad 
	[\hat K_3,\hat K_1] = 4 \hat K_2, \quad
	[\hat K_3,\hat K_2] = 8(\hat K_1^3 - \hat Z \hat K_1).
\ee
The three operators \rf{KK1}--\rf{KK3} separate in the 
coordinate systems introduced in \rf{Syst-I}--\rf{Syst-III}.

\section{Concluding remarks}  \label{sec:seven}

We have introduced the quantum Zernike system 
defined by the Hamiltonian \rf{Zernikeq} that
naturally separates in polar coordinates. This
Hamiltonian is nonstandard because it involves a quadratic
re-scaling potential term, and its wavefunctions have 
nonzero values at its finite circular boundary. 

We have shown that this two-dimensional system
can also be separated in two {\it additional\/} coordinate systems,
where the Zernike Hamiltonian takes the form of quantum
mechanical Schr\"odinger Hamiltonians
with P\"oschl-Teller potentials, whose solutions 
involve separated Legendre and Gegenbauer polynomials.
These coordinate systems become evident when orthogonal
coordinates on a half-sphere are mapped as non-orthogonal
coordinates on the disk.  
The boundary condition on the disk requires one additional 
limit that the solutions on the half-sphere must satisfy.
Associated to the separable coordinate systems, there
are operators whose eigenvalues are constants of the 
motion. Previously only the angular momentum of circular
harmonics was known; this, plus the two new operators
stemming from separability, yielded three operators that
commute with the Zernike Hamiltonian, and close
into a cubic Higgs superalgebra.

We realize that the analysis performed here on the
sphere can be generalized. First, also the 
{\it elliptical\/} system of coordinates
on the sphere and its projections  \cite{PSW1,LS1,L1} 
can be used to separate the Zernike equation and 
provide solutions on one more system of coordinates.
Interbasis expansions will then relate the Zernike functions
on the disk with Legendre and Gegenbauer polynomials
as well as Lam\'e functions. We leave this as a separate 
analysis to be studied elsewhere. We also note that instead
of unit radius and $\alpha_\ssr{Z}=-1$, we may have a
self-adjoint Hamiltonian when the circular boundary 
is at $r=1/\sqrt{\abs\alpha}$, provided that 
$\beta=-2\abs\alpha$. Finally, one can disregard the boundary
problem and revert to the full parameter ranges of 
$\alpha$ and $\beta$, such as was done in Ref.\ \cite{PWY} 
and obtain solutions that correspond to open 
hyperbolic trajectories and, more
generally, study Schr\"odinger equations that stem from
quadratic extensions of the oscillator 
algebra. The methods of solution and mathematical
structure can be along the lines of this research.


\section*{Acknowledgements}

We acknowledge the interest and early discussions with Prof.\ Natig
M.\ Atakishiyev (Instituto de Matem\'aticas, {\sc unam});
we thank Guillermo Kr\"otzsch ({\sc icf-unam}) for indispensable help
with the figures. G.S.P.\ and A.Y.\ thank the support of project 
{\sc pro-sni-2017} (Universidad de Guadalajara). 
C.S.-A. and K.B.W.\ acknowledge the support of {\sc unam-dgapa} 
Project {\it \'Optica Matem\'atica\/} {\sc papiit}-IN101115.


\end{document}